\documentclass{aastex631}

\usepackage{enumitem}

\begin{document}

\title{X-ray and Spectral UV Observations of Periodic Pulsations in a Solar Flare Fan/Looptop}

\author[0000-0001-9726-0738]{Ryan J. French}
\affiliation{National Solar Observatory, 3665 Discovery Drive, Boulder, CO 80303, USA}

\author[0000-0002-6835-2390]{Laura A. Hayes}
\affiliation{European Space Agency, ESTEC, Keplerlaan 1, 2201 AZ Noordwijk, The Netherlands}

\author[0000-0001-8975-7605]{Maria D. Kazachenko}
\affiliation{Laboratory for Atmospheric and Space Physics, University of Colorado Boulder, Boulder, CO 80303, USA}
\affiliation{National Solar Observatory, 3665 Discovery Drive, Boulder, CO 80303, USA}
\affiliation{Department of Astrophysical and Planetary Sciences, University of Colorado Boulder, 2000 Colorado Avenue, Boulder, CO 80305, USA}

\author[0000-0002-6903-6832]{Katharine K. Reeves}
\affiliation{Harvard–Smithsonian Center for Astrophysics, 60 Garden Street, Cambridge, MA 02138, USA}

\author[0000-0002-9258-4490]{Chengcai Shen}
\affiliation{Harvard–Smithsonian Center for Astrophysics, 60 Garden Street, Cambridge, MA 02138, USA}

\author[0000-0002-9690-8456]{Juraj L\"{o}rin\v{c}\'{i}k}
\affiliation{Bay Area Environmental Research Institute, NASA Research Park, Moffett Field, CA 94035-0001, USA}
\affiliation{Lockheed Martin Solar and Astrophysics Laboratory, Building 203, 3251 Hanover Street, Palo Alto, CA 94304, USA}

\begin{abstract}
We present simultaneous X-ray and spectral ultraviolet (UV) observations of strikingly-coherent oscillations in emission from a coronal looptop and fan structure, during the impulsive phase of a long-duration M-class solar flare. The $\approx 50$ s oscillations are observed near in-phase by Solar Orbiter/STIX, GOES, and IRIS Fe XXI intensity, Doppler and non-thermal velocity.
For over 5 minutes of their approximate 35 minute duration, the oscillations are so periodic (2-sigma above the power law background), that they are better described as `periodic pulsations' than the more-widely documented `\textit{quasi}-periodic pulsations' often observed during solar flares. 
By combining time-series analysis of the the multi-instrument datasets with comparison to MHD simulations, we attribute the oscillations to the magnetic tuning fork in the flare looptop-fan region, and betatron acceleration within the lower-altitude flare loops.
These interpretations are possible due to the introduced \textit{Sliding Raster Method} (\textit{SliRM}) for analysis of slit spectrometer (e.g. IRIS) raster data, to increase the temporal cadence of the observations at the expense of spatial information.

\end{abstract}

\keywords{Solar flares --- QPPs --- Oscillations}

\accepted{to ApJ, November 2024}

\section{Introduction} \label{sec:intro}


The standard eruptive model of solar flares \citep[often referred to as the `CHSKP' model after the authors  ][]{Carmichael1964,Hirayama1974,Sturrock1968,Kopp1976}
can account for many of the observed phenomena in solar flares, but there are certain observables that the model cannot explain. Most yet-unexplained behavior relate to higher energy phenomena within flares, including the magnetic reconnection, particle acceleration and transport dynamics within the event. The observational signature of Quasi-Periodic Pulsations (QPPs) are one such example \citep{nakariakov2009, Zimovets2021}.
QPPs are pulsations in the lightcurves of solar flare fluxes and velocities at multiple-wavelengths, showing some characteristic timescale or periodic component. As the name suggests, these pulsations are not perfectly periodic, thus aptly named \textit{quasi}-periodic. QPPs have been observed for many decades, with the first detections reported in \citet{Parks1969,Chiu1970}, in hard X-ray bremsstrahlung and microwave gyrosynchrotron emission, associated with accelerated electrons. 
A few decades later, and thousands of QPP flares have been identified, with periods ranging from seconds to tens of minutes. The majority of these pulsations are reported at wavelengths observed by full disk instruments, including the aforementioned hard X-ray and microwave emission, but also in soft X-ray observations \citep[e.g. those taken by the GOES X-ray sensor,][]{simoes2015, hayes2019, Hayes2020} and radio waves \citep[e.g.][]{Kupriyanova2016,Carley2019}. Quasi-periodic plasma inflows toward the reconnection site have also been observed in extreme ultraviolet (EUV) imaging \citep[e.g.][and references therein.]{hayes2019,Li2020,Mondal2023}.

Due to the limited field-of-view (FOV) of EUV and UV spectroscopic instruments, spectral observations of solar flare QPPs in (E)UV are rarer. That being said, flare oscillations have still been detected in flare ribbons by UV spectroscopy, including Interface Region Imaging Spectrograph \citep[IRIS,][]{DePontieu2014} studies of non-thermal and Doppler velocity oscillations \citep{Brannon2015,Jeffrey2018, Lorincik22}, intensity variations along and across the ribbon propagation \citep[][respectively]{French2021,Naus2022}, and reconnection flux traced by the ribbons \citep{Corchado2024}. The interpretation of on-disk spectroscopic observations of (E)UV QPP sources is difficult, due to the ambiguity of where in the LOS (and thus where in the flare) the oscillations are originating from, and saturation issues during large flares. 

The challenges of interpreting integrated coronal emission along the LOS are fewer in off-limb flares, where higher altitude coronal flare structures are far more easily observed. IRIS Fe XXI spectroscopy of coronal flare structures have also revealed QPPs, including at the base of a plasma sheet (heated plasma around a suspected current sheet), e.g. \citet{Shen2023}, and within Supra-Arcade Fans \citep[hot, wispy, cloud-like structures spanning tens of arcseconds above the flare arcade, typically most visible in soft X-rays and hot EUV channels,][]{McKenzie1999,Reeves2011}, e.g. \citet{Reeves2020}.

The studies noted in this section are a non-exhaustive summary of QPP studies, as QPPs have been detected at almost every wavelength in which we've observed solar flares \cite[see][for reviews on the topic of QPPs in flares]{nakariakov2009, vandoorsselare2016, Kupriyanova2020, Zimovets2021}. QPP-like phenomena have also been observed in stellar flares \citep[e.g.][and references therein]{Mitra2005,Pugh2016,Doyle2018}.
With such a range of observed periods, durations and source locations, many mechanisms for QPPs have been proposed. It is likely that there is no single process responsible for every QPP ever observed, but that many different processes can drive flare oscillations. Most proposed QPP mechanisms fall into one of two camps. The first is that flare oscillations are caused by the time-oscillating nature of the magnetic reconnection and energy release process, and the second is that MHD oscillations are the driver. An overview of a large number of potential QPP mechanisms are outlined in
\citet{McLaughlin2018} and \citet{Zimovets2021}. We discuss the most relevant of these scenarios to our study in the discussion section of this paper.

In this work, we present simultaneous coronal X-ray and UV spectroscopic observations of QPPs detected in a solar flare fan and looptop region on August 29th 2022. Time-series and spectroscopic analysis of the rich dataset, alongside comparison to MHD simulations, allow us determine the likely origin of the QPPs, and provide insights into the dynamics of the flare looptop-fan region.


\section{Observations} \label{sec:Observations}

\begin{figure*}
\centering \includegraphics[width=15.5cm]{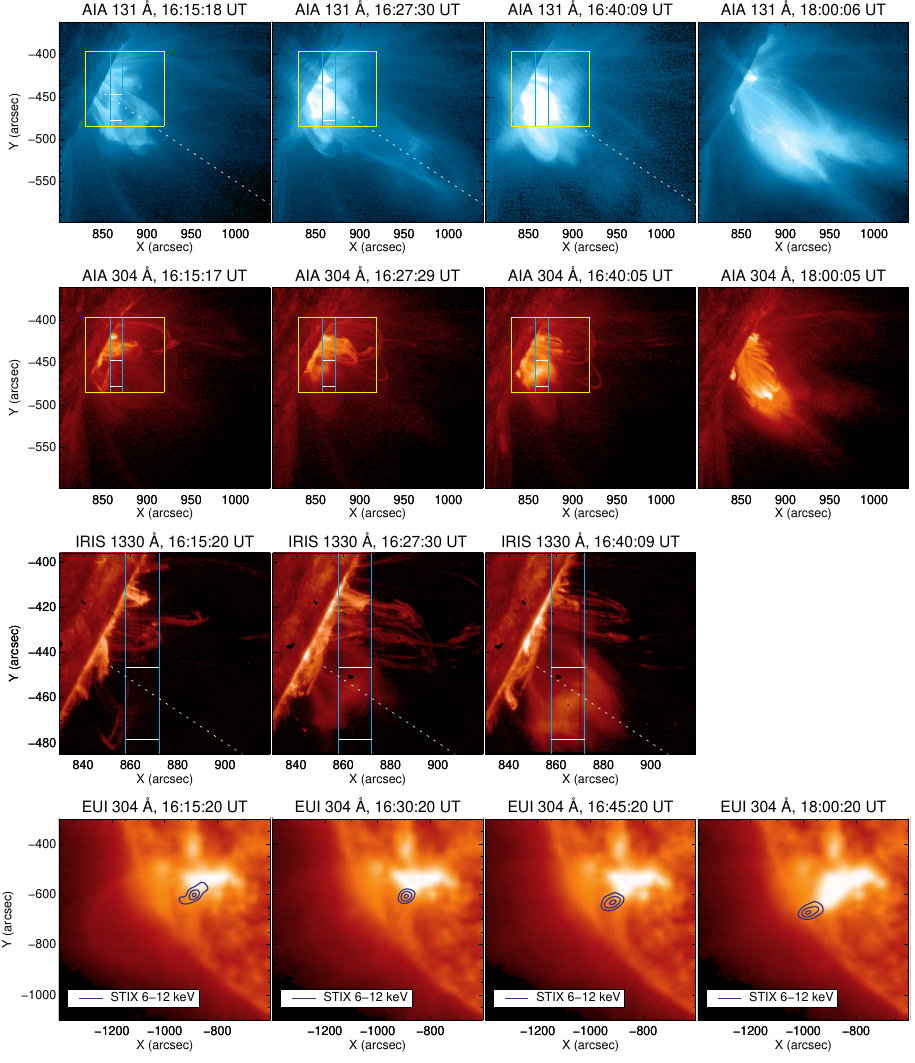}
\caption{Snapshot images of the solar flare evolution. First/second row: AIA 131 and 304 \AA\ respectively. Yellow and blue boxes show the cropped FOV of the IRIS SJI and rastering slit, as shown in row 3. White dashed line shows the cross-section plotted in Figure \ref{fig:ts}. Third row: Cropped IRIS 1330 \AA\ SJI FOV, with blue box showing the east and west edge of the 8 step raster. White dashed line shows the cross-section plotted in Figure \ref{fig:ts}. Bottom row: EUI 304 \AA\ images, taken from a vantage point over the west limb. Blue contours show imaging of 6-12 keV STIX emission. The animated version of the figure show the same four observables evolving with time.}
\label{fig:images}
\end{figure*} 

On August 29th 2022, an M8.6 class flare from AR 13088 was observed simultaneously by instruments at Earth and onboard the Solar Orbiter spacecraft \citep{Muller2020}. From Earth's perspective, the event was seen over the western limb, with the flare ribbons occulted by the solar disk. From Solar Orbiter's vantage point, the flare was observed on-disk, positioned towards the eastern limb. The relative positions of Earth and Solar Orbiter are presented in the right panel of Figure \ref{fig:FFT}. Figure \ref{fig:images} shows near-simultaneous multi-instrument snapshots of the flare's evolution, as observed by the 131 \AA\ and 304 \AA\ channels of the Solar Dynamics Observatory's Atmospheric Imaging Assembly \citep[AIA, ][]{Lemen2012}, 1330 \AA\ slit-jaw imager (SJI) of IRIS, and 304 \AA\ Full-Sun Imager (FSI) observations from the Solar Orbiter Extreme Ultraviolet Imager \citep[EUI, ][]{Rochus2020}. Observations from these instruments are shown in rows 1-4 respectively, at time intervals of 12 minutes for the first three columns of the static figure version, (15 minutes for EUI, due to cadence limitations). The final column shows a flare snapshot later into the event. For this final snapshot, IRIS observations are not available. The animated version of the figure show the observations at greater cadence.

The AIA 131 \AA\ images (Figure \ref{fig:images} - top row) capture emission from Fe VIII and Fe XXI emission in the 0.4 MK (transition region) and 10 MK (hot corona) plasma respectively. AIA 304 \AA\ (Figure \ref{fig:images} - second row) contains contribution from chromospheric and transition region He II emission (sensitive to 50,000 K) and coronal Si XI (1.2 MK) plasma.
The AIA 131 \AA\ images in Figure \ref{fig:images}  show the eruption of a clear flux rope structure (16:15 - 16:27 UT).
Beneath the erupting flux rope, the primary flaring region is saturated (16:40 UT) in 131 \AA, but the 304 \AA\ images show the formation of flare loops. The flare loops continue to grow over the limb for several hours, indicative of the behavior of long duration limb flares previously observed \citep[e.g.][]{French2020, hayes2019}. Above the growing flare loops, from 16:40 to beyond 18:30 UT, we see a high-altitude wispy fan-like structure, prominent in hot AIA 131 \AA\ emission with some lower-altitude signature in cooler 304 \AA\ emission (almost certainly from Si XI emission). This structure resembles the flare Supra-Arcade Fans routinely observed in eruptive limb flares \citep{Reeves2011}. 

The third row of Figure \ref{fig:images} shows observations from the IRIS 1330 \AA\ Slit Jaw Imager (SJI). For context, the FOV of these panels (which have been cropped from the full SJI FOV at the top, left and right edge), are shown as a yellow box in the AIA panels within the same figure. The 1330 \AA\ SJI measures plasma emission from both C II (which probes the chromosphere and transition region at a few 10,000 K), and Fe XXI (flaring corona temperatures of 11 MK). IRIS observed with an 8-step raster, with the east and west bounds of the slit positions marked by the vertical blue lines in rows 1-3 of Figure \ref{fig:images}. Each slit position has a 9.6 second cadence, providing a total raster time of 77 seconds. The SJI captures one image for each slit position. Unlike AIA, IRIS did not capture the flare's full duration. The observing sequence began a few hours before the flare, but ended at 16:48 UT, i.e. $\approx 13$ minutes after the peak of the flare in GOES and IRIS Fe XXI intensity. 
The SJI does not capture the flux rope eruption, but, shortly after the eruption begins in AIA 131 \AA, the SJI sees the formation of distinctive `fuzzy' structure directly beneath the flux rope path, south of the active region loops. This structure is co-located with the saturated AIA 131 \AA\ and unsaturated AIA 304 \AA\ low-altitude fan emission. Fortuitously, the IRIS slit rasters over this structure. 
Examining the IRIS spectra, we find the off-limb structure contains exclusively hot Fe XXI 1354 \AA\ emission, with no trace of the cooler chromospheric C II 1335/1336 spectral lines -- until the very end of the observing sequence, when some cooler loops begin to appear. By tracking the low altitude fan structure in AIA 304 \AA\ observations, (Movie of Figure \ref{fig:images}), we see the plasma grow into the long-lived high-altitude flare fan. Unfortunately, the IRIS data ends before observing the extended fan growth. At the start of the SJI time series, we also see the formation of flare loops at the base of the SJI emission structure, which later reappear in C II as cooler emission. We therefore deduce that the observed SJI emission contains the early stages of hot flare fan region, with hot flare loops convolved into the structure at lower altitudes.  From here on in, we will refer to this region as the flare looptop-fan structure.

\begin{figure*}
\centering \includegraphics[width=15.5cm]{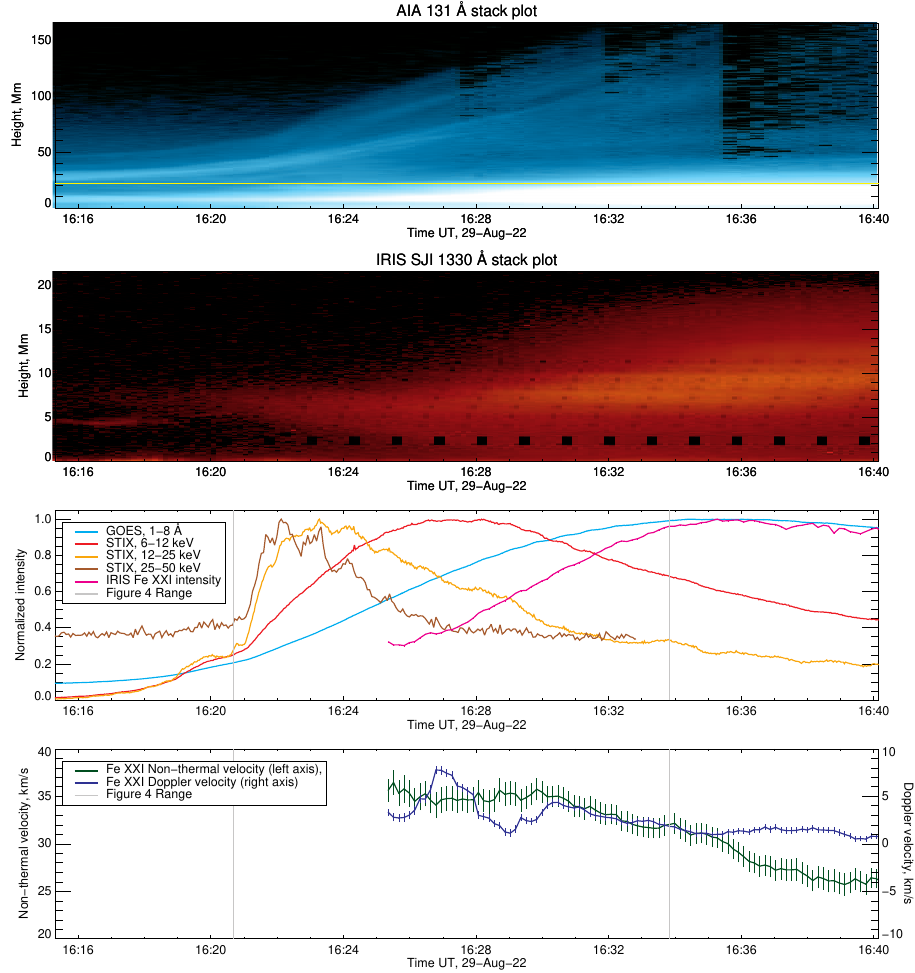}
\caption{Evolution of observed parameters during the flare, with aligned X-axes. Row 1: Flux rope eruption observed by AIA 131 \AA, yellow line shows upper limit of plot in row 2. Row 2: Growth of the fan structure, observed by the IRIS 1330 \AA\ SJI. Diagonal black dots show movement of rastering slit, and larger black squares an artifact of the SJI. Row 3: Time series of GOES 1-8 \AA, STIX 6-12 keV, STIX 12-25 keV, STIX 25-50 keV and IRIS Fe XXI intensity. Row 4: IRIS Fe XXI non-thermal velocity and Doppler velocity. The vertical gray lines in rows 3-4 mark the time period analyzed in this study. }
\label{fig:ts}
\end{figure*} 

The top two rows of Figure \ref{fig:ts} show the height-time profiles for AIA 131 \AA\ and IRIS 1330 \AA\ SJI, located along the dashed lines within the relevant panels of Figure \ref{fig:images}. In the 131 \AA\ plot, we see the eruption of the flux rope, beginning around 16:16 UT. 
Starting around 16:20 UT, we see the growth of the saturated region around 10 Mm above the limb, which expands over the duration of the plot. As this saturation region emerges, we see the response of AIA's variable exposure rate, in the form of black column strips in the height-time plot. The horizontal yellow line in this panel marks the maximum extent of the height-time plot of IRIS SJI 1330 \AA\ (in the second row of Figure~\ref{fig:ts}).
The height-time plot of IRIS SJI 1330 \AA\ (Figure \ref{fig:ts}) reveals the fan beginning to form at the time of flux rope initialization, fully forming before the flux rope leaves the AIA FOV. 
The periodic black squares at 2-3\arcsec\ above the limb are the result of a black speck on the IRIS SJI, which moves with the rastering FOV.
The repeating diagonal pattern of eight dots are caused by the IRIS slit passing over the chosen cross-section location of the height-time plot.

Soft X-ray measurements are obtained by the Geostationary Operational Environmental Satellite (GOES) X-ray sensor (XRS). GOES/XRS observes the full-disk integrated soft X-ray flux from the Sun in the 1-8 and 0.5-4 \AA\ passband channels. We investigate the 1-8 \AA\ emission in this study, but note that both channels show similar behavior during this event. GOES observed with a standard cadence of 1 second. The time series of GOES 1-8 \AA\ emission is plotted in the third panel of Figure \ref{fig:ts}, aligned spatially with the height-time profiles in the previous panels.

From a angular separation of $\approx 150^\circ$, Solar Orbiter observed the flare with the EUI/FSI and Spectrometer Telescope for Imaging X-rays \citep[STIX, ][]{Krucker2020}.
EUI/FSI observed the full disk with two broadband filters, 174 \AA\ and 304 \AA, with a spatial sampling of 4.4\arcsec. Due to the low cadence of 15 minutes, EUI/FSI did not capture much of the CME eruption or flare evolution, but does provide some context images of flare ribbons and subsequent flare loops. The maximum cadence EUI 304 \AA\ images are shown in the bottom row of Figure \ref{fig:images}.
STIX continuously observes the full disk in 4-150 keV energy range, measuring spectra and imaging of both soft and hard X-ray sources. Contours of 6-12 keV X-ray emission are overlaid on the EUI images in Figure \ref{fig:images}, locating the source above the EUV looptops. This location is spatially consistent with the looptop-fan structure observed by the IRIS SJI. 
At 6-12 keV, the X-ray emission is produced primarily by thermal emission from the plasma, observing the hottest flare plasma (most likely newly reconnected loops, at higher altitudes than those typically visible in EUV), with little non-thermal contributions.
The 6-12 keV STIX observations are captured with a cadence of 2 seconds. The time series of STIX 6-12 keV, corrected for light-travel time, is plotted alongside the GOES evolution in the third panel of Figure \ref{fig:ts}. We also plot the time series of 12-25 (thermal and non-thermal) and 25-50 keV (mostly non-thermal) emission in the same panel. The time evolution observed in this plot is discussed within section \ref{ts}.


\section{IRIS spectral analysis} \label{sec:ts}

\begin{figure*}
\centering \includegraphics[width=15cm]{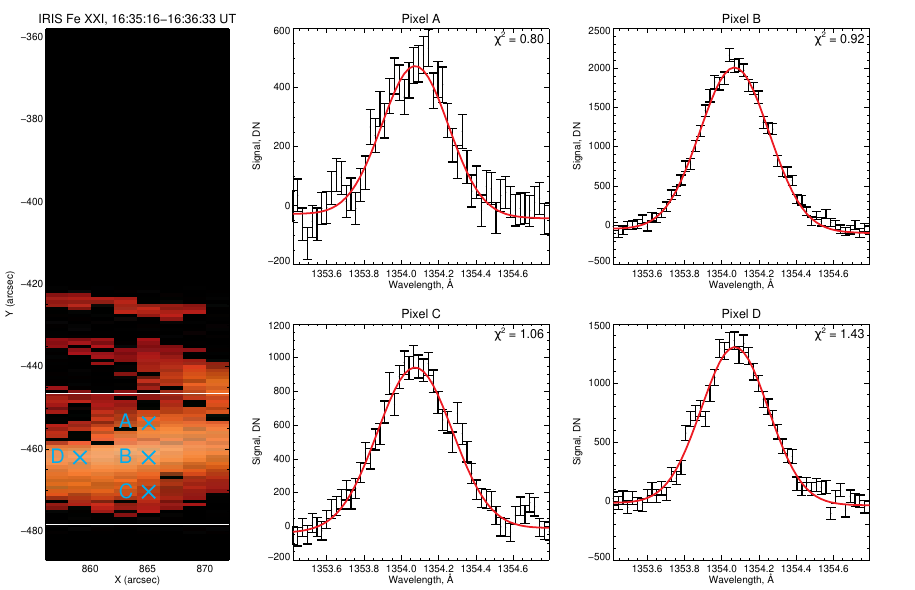}
\caption{Example IRIS Fe XXI 1354.8 \AA\ spectral map, from the peak of Fe XXI intensity emission. Four example pixels are chosen to demonstrate the quality of the single Gaussian fit.}
\label{fig:spectra}
\end{figure*} 

Fe XXI 1354.8 \AA\ is optically thin in the solar corona, and therefore can be fitted by a Gaussian curve. The Gaussian fit provides measurements of intensity, Doppler velocity and non-thermal velocity (excess broadening beyond the thermal and instrumental width, converted into velocity units). For this dataset, we boost the signal-to-noise by binning 5 pixels in the $Y$ direction (retaining the 8 horizontal raster steps). We set thresholds for the intensity, Doppler velocity and non-thermal velocity, to remove pixels with bad fits. These thresholds were set as $< 6000$ DN for intensity, $> 25$ km/s for Doppler velocity and $> 60$ km/s for non-thermal velocity. 
Because all Fe XXI emission originates from the above-limb flaring corona, we cannot calibrate Doppler velocity measurements to a stationary wavelength.
The Doppler velocity measurements therefore show relative amplitude variations, but are not tuned to an absolute value. A precise Doppler calibration is not required for this analysis. Because the contribution function for Fe XXI is narrow around 11MK, the assumption of a consistent thermal width (which is subtracted to calculate the non-thermal velocity) is a standard one for optically thin (E)UV spectroscopy. In Fe XXI, non-thermal velocities are often attributed to unresolved LOS plasma motions, such as those created by waves or turbulence \citep[see e.g.][]{Polito2019, Ashfield2024}.

A sample Fe XXI intensity map, with example (single) Gaussian fits are presented in Figure \ref{fig:spectra}, alongside their $\chi^2$ values. Fe XXI emission is absent across most of the map, present only in the hot looptop-fan region (between the horizontal white lines), and additional hot loops further north. 

\subsection{Red-Blue Asymmetry Analysis}
To justify our use of a single Gaussian fit to the Fe XXI spectra,
we apply Red-Blue Asymmetry (RBA) analysis to the spectral rasters, in a search for systematic line asymmetries across the FOV. The RBAs of the Fe XXI line were obtained by subtracting intensities in the red wing of the line from those in the blue wing, divided by the peak intensity of the profile \citep[see Eq. (1) -- (3) in][]{Polito2019}.

Red and blue wing intensities were integrated in the velocity range of $\pm 50 - 120$\,km\,s$^{-1}$ from the observed line peak. The calculation of RBAs involved an initial subtraction of the FUV continuum, estimated in spectral regions at least $\pm 200$\,km\,s$^{-1}$ ($\approx0.9$\,\AA) away from the peak of the line, well beyond the extent of the broadest Fe XXI line profiles under consideration. Given the location of the emission above the limb, there was little risk of spectral contamination from cooler lines, that we verified by a visual inspection of the profiles. Nevertheless, for the entire duration of the flare that Fe XXI emission is present, we find no evidence of systematic red or blue asymmetries in the rasters.

\subsection{IRIS Sliding Raster Method}

With 8 slit positions of 9.6 s, each 8-step IRIS raster has a cadence of 77 seconds. However, because 77 seconds is longer than the timescales of interest in this flare (i.e. $\sim$~50s, see Section \ref{FFT}), we introduce a new method for maximizing the cadence of spectral observations, at the expense of sacrificing all spatial resolution. We call this novel method the `Sliding Raster Method' (abbreviated to `SliRM'). This method can be applied to spectral raster observations of `oscillating blobs', with the assumption that the plasma source rastered by the slit (in any y-range) is oscillating with a single period and phase. Such assumptions are likely valid for dense rasters, with little spatial separation between either limits of the raster. Returning to the IRIS SJI observations in Figure \ref{fig:images}, we make this assumption for the region bound by the slit raster X-range (the vertical cyan lines) and the manually selected y-range marked by the horizontal white lines. The upper extent of the y-range was selected to avoid the hot loops which enter the spectral window later in the observations. The lower limit was selected by the lower limit of suitable signal-to-noise spectra. 

The Sliding Raster Method (SlIRM) takes the raw time series of intensity, Doppler velocity and non-thermal velocity measured by the slit (regardless of slit position), averaged over the selected y-range. This initial one-dimensional time series will have an artificial periodicity, dependent on the duration and number of time steps in the raster. For example, because the Fe XXI intensity varies across the spectral raster's FOV, the slit will measure an intensity enhancement each time the slit returns to the position of the brightest region. To remove this influence, we calculate a moving average across the slit raw time-series, with the number of data points included in the moving average set to the number of slit positions in the raster. In the case of our observations, with 8 slit steps, a moving average is calculated over pixels $[n-3:n+4]$, with the assigned timestamp of that value set to the mean observation time of these pixels. The resulting time series is no longer sensitive to the raster cadence, enabling us to search for finer changes in the time series.

Representing this in a different way, we can think of a standard raster image created by the 8 slit positions. We take the mean across this FOV (within the selected y-range), to calculate the average measurements for this region. Next, instead of moving to the next raster of 8 slit measurements, we simply replace the data in the first column (first slit position) with the first data column from the next raster. This way, each data point is the average of an 8-step raster image, but with 7 overlapping columns from time step to time step. This method allows us, for targets matching our assumptions, to push the cadence of a rastering observing sequence to the cadence of individual slit positions -- at the expense of losing all spatial resolution in the rastering direction. Figure \ref{fig:slirm} in the appendix demonstrates this process graphically.

The SliRM time series of IRIS Fe XXI intensity, Doppler velocity, and non-thermal velocity are presented in Figure \ref{fig:ts}, aligned temporally with the height-time plots and GOES/STIX time series. Additional text in the appendix further justifies the SliRM method, by comparing the periodicity of the SliRM Fe XXI intensity with the intensity of the SJI images.

\section{Time Series Analysis}\label{ts}

\begin{figure*}
\centering \includegraphics[width=17cm]{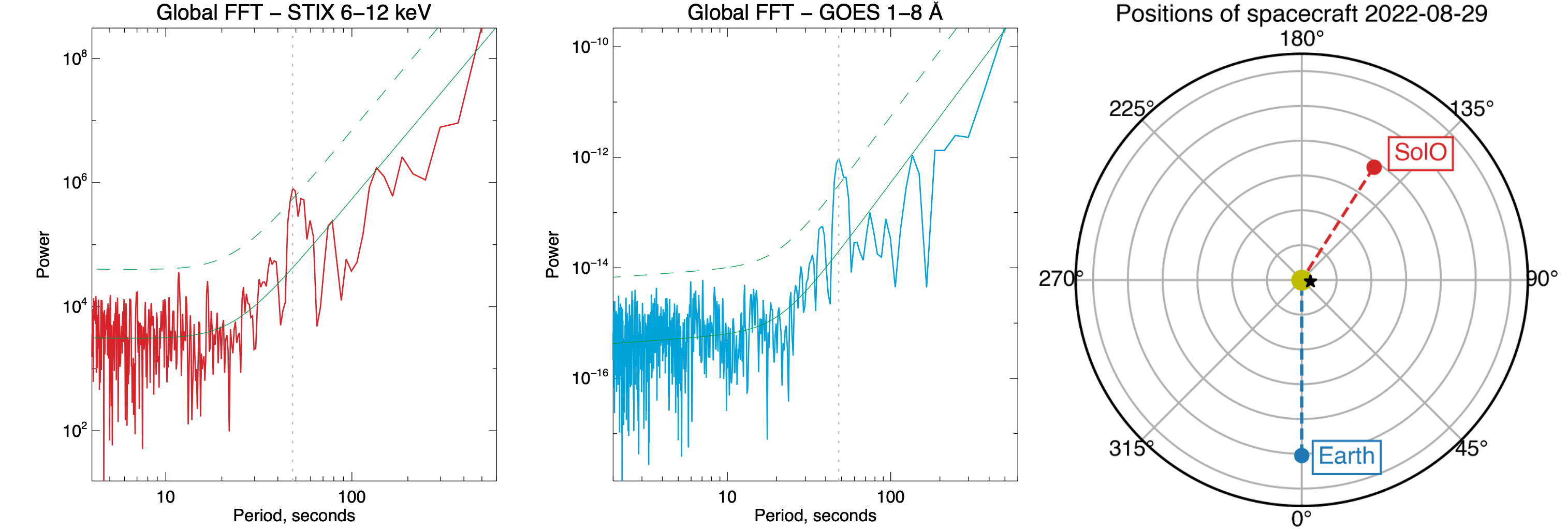}
\caption{Left/Center: Global Fast Fourier Transform (FFT) of raw STIX 6-12 keV (left) and GOES 1-8 \AA\ (center) time series shown in Figure \ref{fig:ts}. The vertical dashed lines mark a local spike at 48 seconds. The solid and dashed green lines show the power law confidence fit and 2 sigma level respectively. Right: Relative locations of Solar Orbiter and Earth observatories (GOES/SDO/IRIS), relative to the Sun.}
\label{fig:FFT}
\end{figure*}

Panel 3 of Figure \ref{fig:ts} shows the time series of STIX 25-50 keV, STIX 12-25 keV, STIX 6-12 keV, GOES 1-8 \AA\ and Fe XXI intensity (in order of most to least energetic plasma).
The highest energy X-rays peak first, and each lower-energy/cooler observables peak progressively later. This is likely due to the hottest flare plasma cooling through the progressively-cooler temperature sensitivity ranges of the multiple instrument channels. 
Of particular interest to this study is the STIX 6-12 keV emission.
From the STIX 6-12 keV images overlaid onto the EUI observations in Figure \ref{fig:images}, we can deduce the 6-12 keV emission originates from hot, dense flare looptop region (emission is dependent on the square of electron density). Given the similar location of the IRIS Fe XXI looptop-fan emission, overlap in the plasma populations observed (from either side of the Sun) by STIX 6-12 keV and IRIS Fe XXI is likely. 

The time series of Fe XXI non-thermal velocity and Doppler velocity (with associated error bars) are presented in the bottom panel of Figure \ref{fig:ts}. 
By the time our Fe XXI observations start ($\approx$ 10 minutes before the intensity peak), the non-thermal velocity is close to its maximum value of 36 km/s. By the end of observations, the non-thermal velocity has decreased to a value of 26 km/s. 
A maximum Doppler velocity of 8 km/s is measured two minutes into the observations, but it is possible that the true maximum value in the source plasma also occurred before our time series begins. It is therefore possible, perhaps even likely, that the Doppler and non-thermal velocities have already decayed from their highest values by the time the Fe XXI intensity is strong enough to be resolved in the observations.

In the STIX 12-25 and 25-50 keV time series (Figure \ref{fig:ts}), QPP-like periodic oscillations are immediately visible from their peak through the decay phase. Of further interest is the STIX 6-12 keV time series, where it is possible to make out more subtle oscillations near the peak of the emission curve. Although less clear that STIX 6-12 keV, GOES also exhibits similarly-subtle small wobbles in the time series during the time period, visible in Figure \ref{fig:ts} to the observant eye. This is true for both GOES channels (but GOES 0.5-4 \AA\ emission is not shown here). From the IRIS time series however, it is trickier to determine from the raw time series whether or not equivalent oscillations are present. The next section aims to pull out relevant oscillatory signals from these time series, by detrending the data.

\subsection{Global Fast Fourier Transform} \label{FFT}

To identify oscillatory signals from the time series, we must detrend the data over a suitable time period. This is required given that the pulsations in the thermal emissions are often low amplitude compared to the overall emission \citep[e.g.][]{simoes2015,Dominique2018,hayes2019, Broomhall2019}. However to first confirm that the detrending method does not introduce artificial oscillations, and to justify which time window to detrend over, we calculate the Fast Fourier Transform (FFT) of the raw (undetrended) STIX and GOES time series between the vertical gray lines in Figure \ref{fig:ts}. These results are presented in Figure \ref{fig:FFT}. Naturally, the power is greater at higher periods (red-noise), due to the fact that flare time series have a power-law shape in the Fourier domain \citep[see][]{inglis2015}. To account for this, we fit and plot a broken power-law to determine the confidence level of the FFT.
In both the STIX 6-12 keV and GOES 1-8 \AA\ power spectra, we note a clear spike at 48 s. 
The spike in the GOES FFT is one of the clearest ever observed in undetrended observations of solar flare QPPs, with the peak exceeding 2 sigma above the power-law background level. The strength of this narrow-band peak in the GOES 1-8 \AA\ global FFT is stronger than the vast majority of the 5519 flare FFTs in the GOES/XRS data examined by \citet{Hayes2020}.
Due to the strong sigma level, the oscillations are better described `periodic pulsations' rather than `\textit{quasi}-periodic pulsations'.

\subsection{Quasi-Periodic Pulsations}

\begin{figure*}
\centering \includegraphics[width=15.5cm]{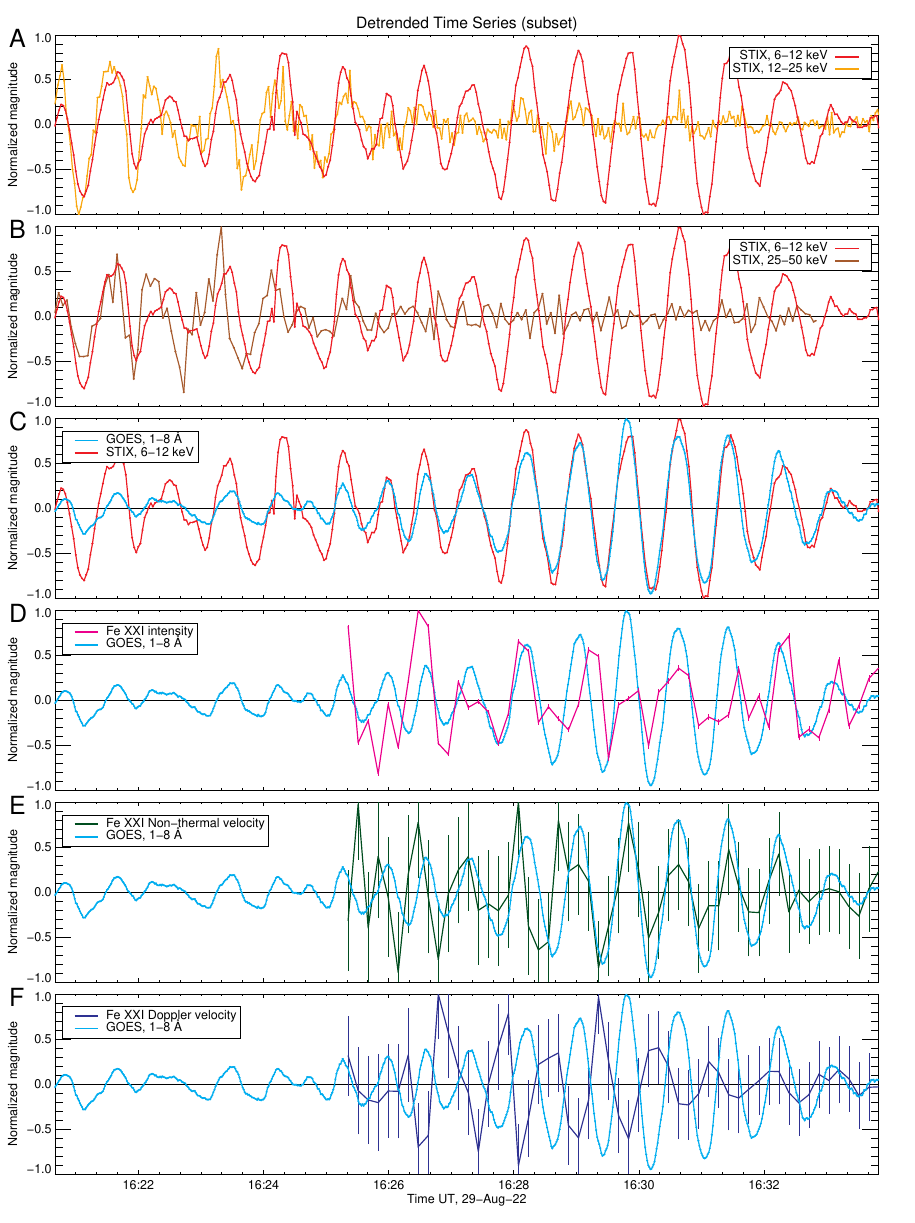}
\caption{Detrended time series over a 100 second period, for the parameters presented in Figure \ref{fig:ts}. The data covers a subset of the Figure 
\ref{fig:ts} time range, spanning the area between the vertical gray lines. The scale of top and middle plots are a percentage of the maximum value. 
}
\label{fig:zoom}
\end{figure*} 

Confident that an oscillatory signal around 50 seconds exists, we can justify detrending the time series of all our parameters with a time scale of 100 seconds, using a Savitzky-Golay filter \citep{savgol_1964, press_1986}. This method is standard for identifying QPPs, as outlined in \citet{Broomhall2019}, and used in other studies \citep[e.g.][]{hayes2019}
The detrended time series are presented in Figure \ref{fig:zoom}, over the time range between the vertical gray lines in Figure \ref{fig:ts}. This is the time range we present for the remainder of this work.
The top three panels of Figure \ref{fig:zoom} compare the detrended STIX 6-12 keV with STIX 12-25, 25-50 keV and GOES emission, whilst the bottom panels compare GOES emission with the IRIS Fe XXI parameters. All time series are normalized by their maximum amplitude. We apply correlation analysis to the oscillations visible across all panels, to determine the phase differences between them (rounded to the coarser-cadence of the two comparison variables).

STIX 12-25 and 25-50 keV (Figure \ref{fig:zoom}A-B) show clearer oscillations at the start of the time series, becoming noisier as they decay with time. These high energy oscillations look reminiscent of typical QPP signatures observed in solar flares. 
Although noisier than the STIX 6-12 keV signal, the higher energy STIX oscillations are close to in-phase with STIX 6-12 keV, preceding by 2 and 4 seconds for STIX 12-25 and 25-50 keV respectively.

Figure \ref{fig:zoom}C, comparing STIX 6-12 keV to GOES 1-8~\AA, is less typical. Between 16:27 to 16:34 UT in particular, we see clear coherent oscillations. These pulsations are particularly striking, with a clear in-phase (0 s) relationship between the two time series observed by STIX and GOES. With a near sinusoidal profile, the coherent oscillations are also the cause of the abnormally strong 2 sigma signal in the global FFT \textit{periodic pulsations}. A detrended time series signal this visually coherent is not standard in QPP observations.
Observed periods are close to 50 s, as suggested by the global FFT.


Further oscillations are found in the time series of IRIS Fe XXI intensity, Doppler and non-thermal velocities. We compare these oscillations to those observed by GOES in Figure \ref{fig:zoom}D-F. Error bars (originating from the Gaussian fit) are plotted for the IRIS values.
Despite the lower cadence and signal-to-noise (particularly in the Fe XXI Doppler and non-thermal velocity data, which result in larger error bars), the Fe XXI oscillations are still consistent with those seen by GOES. For most of the time series, the large error bars are still less than the amplitude of the oscillations.
The Fe XXI Doppler and non-thermal velocities  exhibit their maximum oscillations at the start of the time series, suggesting that their magnitudes are already decaying. The period of the oscillations closely match the period seen with GOES. Rounded to the IRIS cadence of 9.6 s, the Fe XXI non-thermal and Doppler velocities are also in phase (0 s) and anti-phase (28.8 s) to GOES oscillations, respectively. For the Doppler velocity measurements in particular, the qualitative anti-phase signal with GOES is striking. We recall that positive Doppler velocity measurements represent red shift, with plasma moving away from us, with negative velocity representing plasma moving towards us. The direction of the moving plasma is dependent on the position of the plasma source at the limb, and thus the difference between phase/anti-phase in this case is somewhat arbitrary. Doppler velocity oscillations decay quickly, but still show a periodic nature at the end of the time series. The non-thermal velocity measurements are slightly noisier, but still show a close to in-phase relationship with the GOES measurements. 
Finally, Fe XXI intensity measurements show a similar story, with slightly noisier observations. Oscillations are in-phase (0 s) with GOES measurements (considering that the IRIS measurements have a cruder cadence of 9.6 seconds). The peak amplitude of the Fe XXI intensity oscillations does not occur until later in the flare evolution.
To test whether the Fe XXI oscillations are present over the full highlighted y-range, we test the application of SliRM to subsets (in y dimension) of the selected region of interest. We find that although oscillations become noisier with less selected pixels, their period remains consistent over the region. This reveals that the fan and looptops present within this structure are oscillating together, and confirms that the periods are not artifacts of SliRM (as they exist irrespective of the size of the averaging window).

\begin{figure*}
\centering \includegraphics[width=15.5cm]{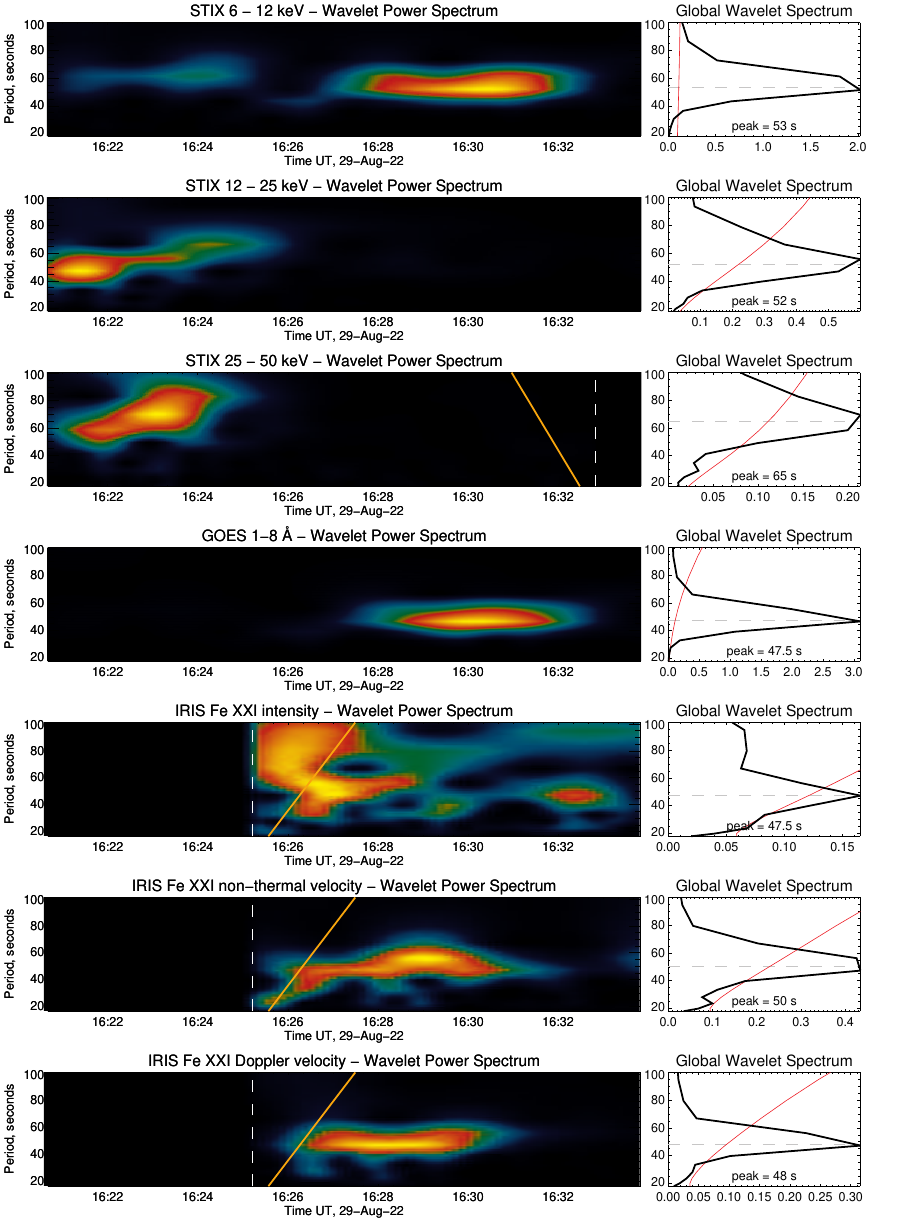}
\caption{Wavelet analysis (Morlet) of the full time-series range plotted in Figure \ref{fig:ts}, cropped to Figure \ref{fig:zoom} time-range of interest for comparison. The vertical dashed lines in some wavelet plots show the start/end time of those observations, with the plot extended to show ease of comparison to other parameters (the Wavelet analysis is applied to the non-extended time series). The orange lines show the cone of influence for the wavelet. For some parameters, the cone of influence boundaries sit outside of the time-range plotted here (but the data is within the cone). In the right column we show the global wavelet spectrum, with red line showing the 95\% confidence interval.}
\label{fig:wavelet}
\end{figure*} 

We also apply wavelet analysis \citep[Mortlet method,][]{torrence_compo} to the time series plotted in Figure \ref{fig:zoom} for each parameter. We apply the wavelet analysis to the detrended data across the full time series (the time range shown in Figure \ref{fig:ts}), but present only our (Figure \ref{fig:zoom}) region of interest in Figure \ref{fig:wavelet}. For the STIX 25-50 keV and IRIS parameters, the time series begin or end within this range. For these values, we can only apply the analysis to the available data range, but still present all parameters on a consistent time scale. 
We also provide the global wavelet and cone of influence for each parameter, although the cropped time-range means the cone of influence boundaries sometimes sit outside our plot region. In these cases, all plotted data is within the cone of influence.

The results of the quantitative wavelet and correlation analysis (discussed above) are summarized below, providing the oscillation period and phase difference relative to GOES 1-8 \AA\ emission. In each case, the phase difference is rounded to the cadence of the coarser-cadence variable (2 s for STIX, 9.6 s for IRIS).

\begin{itemize}
[itemsep=0ex]
    \item STIX 6 - 12 keV: 53 s, 0 s phase difference (in-phase)
    \item STIX 12 - 25 keV: 52 s, 2 s ahead (near-in-phase)
    \item STIX 25 - 50 keV: 65 s, 4 s ahead (near-in-phase)
    \item GOES 1-8 \AA: 47.5 s, (reference for phase correlation)
    \item IRIS Fe XXI intensity: 47.5 s, 0 s phase difference (in-phase)
    \item IRIS Fe XXI non-thermal velocity: 47.5 s, 0 s phase difference (in-phase)
    \item IRIS Fe XXI Doppler velocity: 50 s, 28.8 s phase difference, (anti-phase)

\end{itemize}

All these periods provide a global period of 47.5 -- 53 s, with the exception of the noisier (and shorter duration) STIX 25-50 keV oscillations, which give a longer global wavelet period of 65 s. 
All regions of significant wavelet power fall within the cone of influence, as do the global wavelet peaks.
Also significant is the \textit{lack} of power at a period of 77 seconds within the IRIS Fe XXI parameters. If you recall, 77 s is the cadence of each 8-step raster in the IRIS data, from which we pulled out the 9.6 s cadence time series using the SliRM method. If we were to see a signal at 77 seconds, it would show the SliRM to be unsuccessful, by polluting the temporal time series with spatial variations across the raster FOV. We also note that the wavelet analysis reveals no notable period drift across the observation period,

\section{Fan Oscillations in Flare Simulations} \label{sec:sim}

\begin{figure*}
\centering \includegraphics[width=15.5cm]{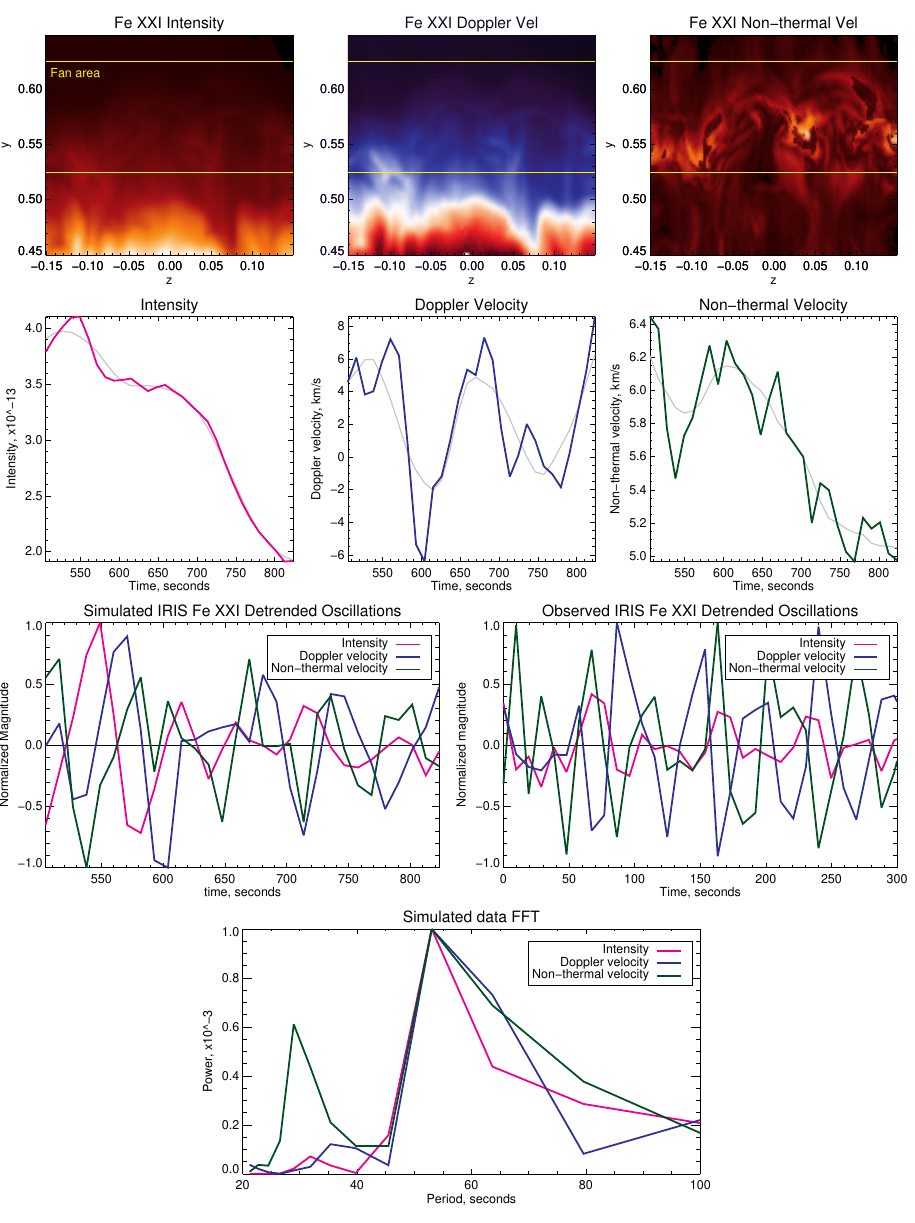}
\caption{Top row: LOS integrated maps of simulated Fe XXI parameters. Second row: Time series of mean emission over the yellow box in above row, with smoothed curve overplot in gray. Third row: Comparison between detrended oscillations in both simulated and observed Fe XXI emission. Bottom row: FFT of the detrended simulated Fe XXI time series.}
\label{fig:simulation}
\end{figure*} 

To better interpret the time series analysis of the IRIS time series data, we draw comparisons between the observed IRIS oscillations and oscillations present within a three-dimensional flare model \citep{Shen2022}. In this model, we perform the resistive magnetohydrodynamic (MHD) simulation to model magnetic reconnection in the standard solar flare geometry and reveal highly dynamic plasma flows in the reconnection current sheet and flare loop-top regions. This simulation is initialized based on the typical two-ribbon flare scale, which was not customized to the parameters of the August 29th 2022 flare specifically. As demonstrated in \citet{Shen2023}, IRIS Fe XXI emission can be synthesized by integrating emission along the LOS from any direction. The top row of Figure \ref{fig:simulation} shows the synthesized Fe XXI intensity, Doppler velocity and non-thermal velocity for an example frame of the simulation run. Our vantage point is perpendicular to the flare arcade, with the brighter flare loops visible below $y=0.5$ in the intensity frame. Above the flare loops, we see the flare fan structure, subtle in intensity and Doppler velocity, but enhanced and prominent within the non-thermal velocity map. Other than the magnitude of the observed parameters, there is no key difference between the flare fan and loop regions within the simulation. In terms of magnetic connectivity, they are indistinguishable, with both regions found to oscillate and evolve together.

Given the use of our SlIRM routine with the IRIS observations, our observed Fe XXI oscillations are averaged over the global fan structure, and not unique to a single point in the plane-of-sky (POS). To replicate this in the simulations, we take a mean of the Fe XXI parameters over the entire global fading (or fan) structure above the most bright flare loops regions, marked with the yellow box in Figure \ref{fig:simulation}. The resulting time series is plotted in the second row of Figure \ref{fig:simulation}. Intensity and non-thermal velocity show a downwards trend across the time series, whereas the Doppler velocity shows oscillations towards and away from us. To search for oscillations of the scale found in our observations (10s of seconds), we apply our previous detrending and FFT analysis (applied to observations in Section \ref{sec:ts}). The gray line alongside the time series in Figure \ref{fig:simulation} shows the moving average of the data, which we subtract to calculate our detrended variables in the first panel of the Figure \ref{fig:simulation} third row. The final row calculates the FFT of the oscillations, for both the detrended and raw time series. The FFT reveals oscillations of 53 seconds in the simulation, present simultaneously across Fe XXI intensity, Doppler and non-thermal velocity. 

The period of Fe XXI oscillations in the simulation remarkably match the period detected in our observations ($\sim 50$s). Even more remarkably, the magnitude of the intensity and non-thermal oscillations (not depicted in the graph due to the normalized magnitude), also match the observations. The Doppler velocity oscillations, however, are about an order of magnitude larger than observed. This result could partly be explained by the difference in viewing geometry between the simulation and observations. As previously mentioned, this simulation run was not customized to the parameters of the August 29th 2022 flare specifically, so the fact we see similar oscillation magnitudes and periods is somewhat fortuitous, as many of these variables can be adjusted through the change of simulation characteristic scales. What is important, however, is the confirmation that oscillations of \textit{equal} periods are found between the three IRIS parameters -- a critical constraint from our observations. The simulation also replicates the anti-phase between Fe XXI intensity and Doppler velocity, with a phase difference of 26 s. It does not, however, replicate the phase difference to non-thermal velocity. This is potentially due to asymmetries of structures in the POS and LOS of the simulation, which are not replicated in reality.

\section{Discussion} \label{sec:disc}

In this study we have identified highly coherent QPPs in a solar flare, identifying the source region as tenuous emission at and above the growing flare looptop-fan. This tenuous emission later goes on to become a large fan structure, inclusive of faint Supra-Arcade Downflows \citep[SADs, see e.g.][]{Savage2011,Xie2022}. The oscillations are only observed in the hottest channels.
Utilizing simultaneous observations of this region with Fe XXI spectra from IRIS, and multi-energy-range X-ray observations from STIX and GOES, we have firm observational constraints of the QPP mechanism driving oscillations in the hottest flare plasma, ruling out the majority of candidate processes as the origin. 
The list below summarizes the observational constraints.

\begin{enumerate}[itemsep=0ex]
    \item Oscillations of period 47.5 - 52 s detected in GOES 1-8~\AA, STIX 6-12 keV, STIX 12-25 keV, and IRIS Fe XXI intensity, Doppler and non-thermal velocity. The noisier STIX 25-50 keV emission reveals a longer period of 65 s.
    \item Peak to trough oscillations of $\approx 1.8$ and $\approx 1.4$ km/s for Fe XXI Doppler and non-thermal velocity respectively.
    
    \item STIX 6-12 keV, GOES 1-8~\AA\ and Fe XXI parameters are in phase with one-another, (recalling that the sign of the Doppler velocity is arbitrary, relating to the location of the plasma in the LOS), within the IRIS cadence of 9.6 s. The STIX 12-25 and 25-50 keV emission are ahead by 2 and 4 s respectively.
    
    \item No period drift throughout the 36+ minute observation window (so no correlation to loop-length, which grow throughout the flare).
    \item STIX 6-12 keV emission is located at the hot flare looptop, with a tenuous IRIS Fe XXI source likely containing emission from both the looptop and fan. This Fe XXI source begins growing as the flux rope erupts. As far as we can tell from the different vantage points, these two regions are located close to co-spatially.
    \item A recent 3D MHD simulation show similar oscillations in the global fan structure, with matching periods between Fe XXI intensity, Doppler and non-thermal velocity.
    \item No systematic red-blue asymmetry in the Fe XXI spectral profiles.
    \item The QPPs are observed simultaneously from two vantage points $\approx 150^\circ$ apart. This is consistent with an optically-thin QPP source in the corona. 
\end{enumerate}

With the above observational constraints, we can start to rule out potential popular QPP mechanisms. In the literature, QPP mechanisms primarily fall into two camps -- MHD oscillations and bursty/periodic reconnection. We discuss the (im)possibility of mechanisms within these two categories below.

The first thing to consider is that the GOES and STIX 6-12 keV oscillations, which are in phase during the 16:27 -- 16:33 UT window, are observing the flare from near-opposite sides of the Sun. During this time period, the time series observed by each instrument are strikingly coherent and periodic ($>$2 sigma confidence), more accurately described by `periodic pulsations' than `quasi-periodic pulsations'. The coherence between signals from these two instruments further suggest they're observing the same pulsating plasma population. However, because the pulsations are only a few percent of the overall SXR flux, the modulations themselves could still be originating from a relatively small volume, and not necessarily from within a large coronal structure.

Another constraint for a large range of potential QPP candidates was the lack of period drift observed across the $\sim$ 36 minutes of observations. This result is shown within our Wavelet analysis in Figure \ref{fig:wavelet}. We suspect the Fe XXI emission observed by IRIS to contain both a fan structure and flare loops. In AIA, we see significant growth of the flare loop arcade throughout the 36 minutes of IRIS observations (Figure \ref{fig:images}). Therefore, any QPP mechanisms relating to oscillations of flare loops, with a period relating to the length of the loop, are very unlikely. This observational constraint rules out the role of magneto-acoustic slow-mode waves in this event, e.g. \citet{Wang2002, Wang2015,Pant2017}.
It is important to note, however, that although the overall flare loop arcade grows during a solar flare, individual loops do not. It could therefore be possible that a mechanism within an individual set of hot loops at a given height is driving the long duration oscillations, whilst the same mechanism is not acting on subsequently-formed loops at higher altitudes.

Despite a dependency on loop-length, previous observations of oscillations, determined to be driven via the global sausage mode, bare many similarities with our observations of this flare. \citet{Tian2016} present IRIS and GOES observations of oscillations in an on-disk M1.6 class flare, with a period of 25s. This period is only a factor of two less than our observations. The authors also find matching periods between GOES, Fe XXI intensity and Doppler velocity, as we do here. \citet{Tian2016} prescribe these observables to the global sausage mode, presenting convincing models of the mechanism. 
However, a key prediction of this model is a period of Fe XXI non-thermal velocity half that of the Doppler and intensity periods. The IRIS observations in \citet{Tian2016} do not have the cadence necessary to validate this prediction, but with a longer oscillation period present in the August 29th 2022 event studied in this work, this lower non-thermal velocity period would be observable. Despite the large error bars in the Fe XXI non-thermal velocity (Figure \ref{fig:zoom}), the error bars are small enough that the peaks and troughs of most of our oscillations are still resolvable. Crucially, they do not reveal a period half of the other Fe XXI parameters, ruling out the role of the global sausage mode. Furthermore, the sausage mode predicts a phase shift between intensity and Doppler velocity, which we similarly don't observe in our data. Previous observations and modeling of sausage mode oscillations also typically have far shorter periods than we observe in this study, of order 1-10 s \citep[e.g.][]{Nakariakov2012,Nakariakov2018}.

Given the striking coherence  of oscillations observed in the hottest plasma by GOES and STIX 6-12 keV, especially within the 16:27 -- 16:34 UT window, it is difficult intuitively to imagine a mechanism driven \textit{directly} by periodically-bursty reconnection or reconnection outflows. As our QPP source region is located within the flare loptoop-fan structure, ruling out reconnection-driven QPP mechanisms will provide insight not only into the nature of QPPs in this flare, but also into the dynamics of fan structures (and whether they are related to reconnection outflows).
There are no SADs visible at this stage of the flare, but they become visible in the same off-limb structure during the later phases of the flare. Therefore, the absence of reconnection-driven QPPs in the flare fan (the medium through which SADs pass through), may support or rebut certain SAD models. QPPs have previously been detected in large-scale SADs (as a result of heating), but on far longer timescales (and with far less coherence) than the QPPs we investigate here.
A key prediction of reconnection outflows are the asymmetry of spectral lines \citep{DePontieu2010, Polito2018}. Our Red-Blue Asymmetry analysis of the Fe XXI emission in the fan structure finds no evidence of systematic asymmetry (above normal noise levels), revealing that strong reconnection outflows within the region are unlikely. This result suggests that the fan structure does not directly contain plasma outflows from the reconnecting current sheet, and that the QPP mechanism is more likely linked to some kind of MHD oscillation. The former insight provides evidence to the model that flare fans, and the SADs that sometimes flow through them, are not directly connected to fast current sheet outflows, but instead
are plasma populations beneath a possible higher-altitude termination shock. The lack of reconnection sites in the observed plasma region also rules out the possibly of any myriad of direct current sheet oscillation, e.g. those examined in \citet{Li2016,Shen2023}. Magnetic reconnection must therefore be taking place at an even higher altitude, perhaps above a termination shock.

Another proposed mechanism to drive solar flare QPPs is the 'magnetic tuning fork' oscillation, in an \citet{Takasao2016}-style flare fan \citep[see also ][]{Shibata2023}. Using 2D simulations, \citet{Takasao2016} make the case for oscillations triggered by the \textit{indirect} backflow of reconnection outflows in the flare looptop-fan (remember, there is little difference between the two in simulations). It is worth distinguishing that the oscillations do not take place within the plasma outflows themselves, which we have ruled out as a possibility here. The simulation periods of 10s -- 100s of seconds, dependent primarily on the plasma beta in the region. Figure 4 of \citet{Takasao2016} depicts a region undergoing periodic compression of the above-loop top flare fan. As the plasma is compressed, we would expect an in-phase increase in temperature and density (resulting in the enhancement of IRIS Fe XXI intensity and soft/thermal hard X-ray emission). This is what we observe in our event. Our observed Fe XXI Doppler velocities also fit into this scenario, with a sign of the observed downflow-direction motion dependent on the flare position either side of the limb. The observed Fe XXI non-thermal velocity is interesting, as the pattern would suggest an increase in the non-thermal velocity under compression, indicating that plasma compression is enhancing turbulent or unresolved flows in the LOS. If our Fe XXI and thermal X-ray observations were indeed caused by this scenario, the primary reconnection site is not visible in our observations -- instead, located at a higher altitude than the oscillating looptop-fan region we spatially resolve.
The downwards reconnection outflows then impact the top of the flare looptop-fan, and trigger oscillations within the region.

Further convincing evidence of magnetic tuning fork oscillations have been found in the past, most notably by \citet{Reeves2020} during a flare fan structure above the southern arcade of the September 10th 2017 X8.2-class solar flare. These oscillation periods were nearly an order of magnitude greater than ours, with a period of 400s. The magnetic tuning fork is also present in the 3D simulations examined in Section \ref{sec:sim}, albeit with less symmetry than in 2D work. The plasma beta versus period trend found in 2D models of the tuning fork provide a good reference range, but 3D simulations show the period also depends on the asymmetry of the local environment. This would explain the difference in periodicity found between the September 10th 2017 flare, and August 29th 2022 flare examined in this work.

The only observable criteria not immediately explained by the magnetic tuning fork model are the oscillations of the high energy (with large non-thermal component) 12-25 and 25-50 keV x-rays observed by STIX. At first, the oscillations in non-thermal electrons might suggest inherent periodicity in the energy release process. However, as we have already discussed, given the other observables, reconnection-driven QPP mechanisms are unlikely. Instead, betatron acceleration is able to explain the higher energy electrons, whilst still remaining consistent with the magnetic tuning fork model. Betatron acceleration, first introduced by \citet{Brown1975}, is the acceleration of electrons via the magnetic mirroring effect, allowing electrons to gyrate down the solar surface as the cross-section of the magnetic loops periodically increase (and field strength decreases). As the flux of downflowing electron populations increase, we see an increase in hard X-ray emission. During the magnetic tuning fork oscillations, periodic compression and relaxation of the fan structure above the loops may produce such an oscillation in betatron acceleration.

\section{Summary} \label{sec:concl}

In this work we present analysis of a well-observed, uniquely behaving M8.6-class eruptive solar flare from August 29th 2022. The flare produces extremely coherent periodic X-ray oscillations (GOES and STIX 6-12 keV), resulting in an unusually strong spike ($>2$ sigma confidence) in the global FFT of GOES emission. During this time period, these oscillations are better defined as `periodic pulsations', rather than `\textit{quasi}-periodic pulsations'. The event is well observed by IRIS at Earth orbit, and Solar Orbiter from behind the Sun, with in phase GOES, IRIS Fe XXI and STIX emission. 
To our knowledge, these are the first simultaneous HXR and UV spectral observations of a coronal QPP source off the limb, spatially identifying a hot looptop-fan structure as the origin. With our well defined observational constraints, we determine that magnetic tuning fork oscillations within a \citet{Takasao2016}-style fan is the most likely explanation for the 50-second QPPs, with periodic higher energy X-rays resulting from betatron acceleration. The tuning fork oscillations are also present in 3D flare simulations. The new SliRM method of analyzing spectra from a rastering slit spectrograph made the analysis of IRIS Fe XXI emission (necessary to provide the observational constraints) possible. SliRM opens up many possibilities for future time series analysis of solar data, including, but not limited to, studies of QPPs in solar flares. SliRM should also be applicable to the analysis of data from upcoming slit spectrograph missions, such as NASA's Multi-slit Solar Explorer (MUSE).

\begin{acknowledgments}
\section*{Acknowledgments}
IRIS is a NASA small explorer mission developed and operated by LMSAL with mission operations executed at NASA Ames Research Center and major contributions to downlink communications funded by ESA and the Norwegian Space Centre. Solar Orbiter is a mission of international cooperation between ESA and NASA, operated by ESA.
We thank Hugh Hudson, Lyndsay Fletcher and Sarah Matthews for their useful discussions.
R.J.F thanks the Brinson Foundation for their support via the Brinson Prize Fellowship. J.L. acknowledges support from NASA under contract NNG09FA40C (IRIS) and from NASA H-GI (Open) program grant 80NSSC20K0716. K.K.R. is supported by contract 8100002705 from Lockheed-Martin to SAO. M.K.D. acknowledges support from NASA ECIP NNH18ZDA001N and and NSF CAREER SPVKK1RC2MZ3  awards. C.S. is supported by grants NSF AST2108438 and NASA 80NSSC21K2044 to SAO.  L.A.H is supported by an ESA Research Fellowship. 
\end{acknowledgments}

\bibliography{bibliography}{}
\bibliographystyle{aasjournal}

\begin{appendix}


\section{Justifying SliRM with IRIS SJI data}

\begin{figure*}
\centering \includegraphics[width=12cm]{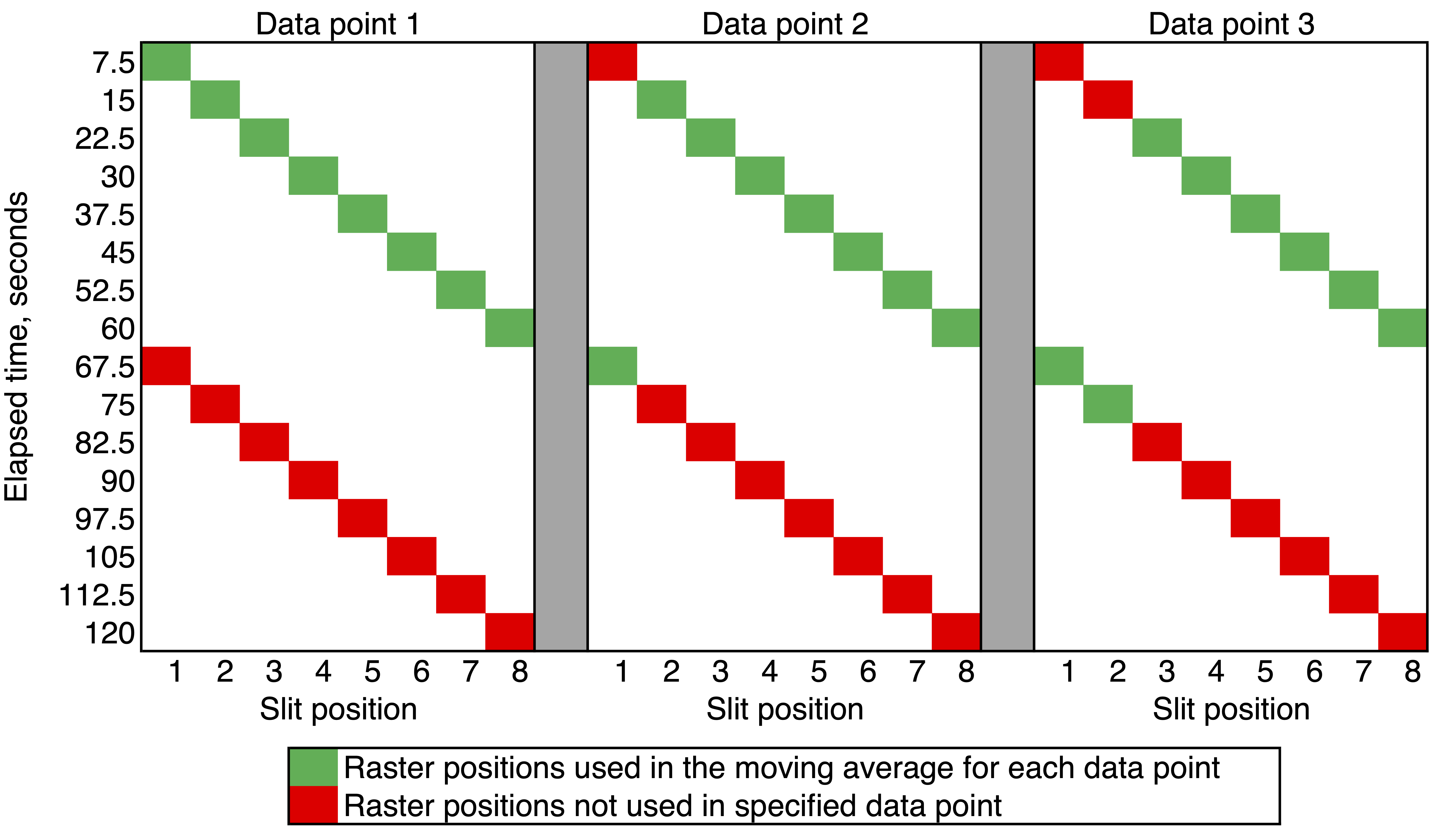}
\caption{Schematic showing the data points used in the SlIRM moving average for the first 3 time steps of an 8-step raster.}
\label{fig:slirm}
\end{figure*} 

\begin{figure*}
\centering \includegraphics[width=15cm]{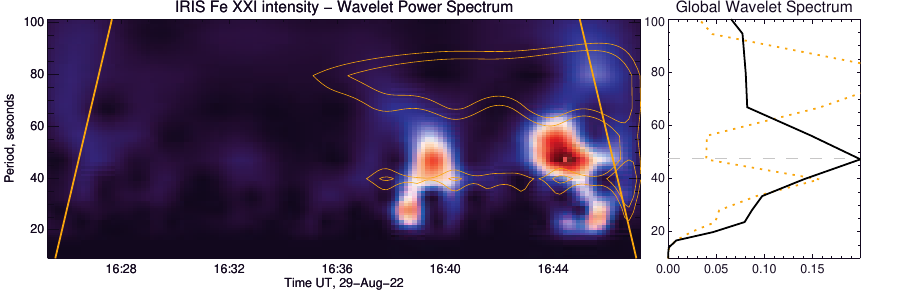}
\caption{Subsection of Figure \ref{fig:wavelet}, showing Fe XXI intensity wavelet analysis. 1330 \AA\ SJI wavelet is overplot in orange.}
\label{fig:FeXXI_wavelet}
\end{figure*} 

The SliRM method aims to extract a time series from a time-space convolved raster, by sacrificing any spatial resolution. In the case of the IRIS raster analyzed in this study, the 77 s raster consists of 8 slit positions with 9.6 s cadence each. Our detrending and wavelet analysis successfully identifies a $\approx$50 s QPP cadence, and crucially detects no signal at 77 s (the length of the raster). An identification of a 77 s period would mean that the SliRM time series is still capturing spatial information, and thus unsuccessful. To further validate the use of SliRM on these observations, we can compare the wavelet analysis of the SliRM Fe XXI intensity time series, with that of the SJI observations averaged over the same region. We can do this because the SJI 1330 \AA\ signal in this off-limb region is exclusively from Fe XXI emission. Figure \ref{fig:FeXXI_wavelet} shows this comparison. The colorized image shows the wavelet of SliRM Fe XXI intensity, and orange contour the overlay of the SJI wavelet. The adjacent panel to the right compares the global wavelet of both. As discussed previously, the SliRM Fe XXI intensity is dominated by the 47.5 s signal. The SJI wavelet, however, is dominated by two peaks. The first is found at 77 s, equal to the raster cadence, which we expect to see in this time series due to the slit motion across the SJI. The second period is detected at 40 s. This signal is lower than the 47.5 s signal found in the Fe XXI SliRM wavelet, albeit by less than one time step (9.6 s). Given the low 9.6 second cadence (compared to the period we are searching for), both methods are likely measuring the same period in the pulsations, validating the use of SliRM on the Fe XXI spectra.


\end{appendix}

\end{document}